\documentclass[aps,prl,twocolumn,groupedaddress,showpacs]{revtex4}
\usepackage{graphicx}

\begin{document}


\title{Current driven magnetization dynamics 
in helical spin density waves}


\author{Ola Wessely}
\affiliation{Department of Physics, Uppsala University,
Box 530, SE-75121, Uppsala, Sweden}
\author{Bj\"orn Skubic}
\affiliation{Department of Physics, Uppsala University,
Box 530, SE-75121, Uppsala, Sweden}
\author{Lars Nordstr\"om}
\affiliation{Department of Physics, Uppsala University,
Box 530, SE-75121, Uppsala, Sweden}


\date{\today}

\begin{abstract}
A mechanism  is proposed for manipulating the magnetic state of a
helical spin density wave using a current. In this paper,
we show that a current through a bulk system with a helical 
spin density wave induces a spin transfer torque,
giving rise to a rotation of the order parameter.
The use of spin transfer torque to manipulate the magnetization in bulk 
systems does not suffer from the 
obstacles seen for magnetization reversal using interface spin transfer torque
in multilayered systems. 
We demonstrate the effect by a
quantitative calculation of the current induced magnetization dynamics
of Erbium. Finally we propose a setup for experimental verification.

\end{abstract}



\maketitle

The possibility to manipulate the magnetization of a magnetic material 
using the spin transfer torque (STT) was proposed from theoretical 
considerations ten years ago by
Slonczewski \cite{Slon} and Berger \cite {Berg}. Since then STT has been
a very active area of research boosted by its potential technological 
application in magnetic random access memories.\cite{Fert} 
The phenomenon has been studied in layered materials where it 
is considered to be an interface effect occurring when a current 
passes an interface between two regions with different magnetization 
direction.\cite{Stiles} 
Experimentally, it has been shown that the 
current induced torque 
is able to switch the magnetization
of a ferromagnetic layer in a nano pillar containing
a ferromagnetic/non magnetic/ferromagnetic 
trilayer.\cite{Kat}
An obstacle for using the STT in technological applications is the 
magnitude of the smallest
current needed to switch the magnetization. 
Theoretical models of trilayer systems 
have shown that 
the size of 
this critical current is determined by the magnetic anisotropies.\cite{Stiles2} 
%
%
%

In this work we consider materials with a helical spin density wave (SDW),
a magnetic ordering which mathematically can be described as a spin
spiral (SS).\cite{Herring,Sand} 
In a SS 
the direction of the magnetization rotates as one moves along 
the SS wave vector. Without loss of generality we 
consider the magnetization rotation axis to be parallel 
to the 
SS wave vector
and will refer to it as the spin spiral axis.
In comparison with a multilayer a SS 
can
be considered as a self assembled
magnetic multilayer system where the atomic layers 
perpendicular to the spiral axis are ferromagnetic,
but where the magnetization of the different layers 
are non parallel with respect to each other.
We predict that a new kind of STT will be induced by a current parallel
with the spiral axis. This phenomenon 
is a bulk effect in contrast to the STT in a multilayered 
nano pillar. An advantage with a bulk system 
is the possibility to better 
control the magnetic anisotropies that interfere with the STT in an 
artificial multilayer.

First we give a general argument for the  occurrence of STT in SS systems and then
we establish the effect by a  
quantitative calculation
of the current induced STT in Er. Er is chosen since it is one of the many
well known examples of a helical SDW in the rare earth (RE) series
(see e.g.\ Ref. \onlinecite{Nord}). 
The calculation is performed 
by combining first principles calculations
and semiclassical linear response theory.

The STT in multilayered systems can be understood by considering
a trilayer system A/B/C of two ferromagnetic layers A and C,
separated by a non magnetic layer B.
If the two ferromagnetic layers 
A and C have non parallel magnetization, a current passing 
perpendicular to the layers,
from layer A into 
layer C will change its spin polarization when passing into
layer C.\cite{Stiles} The
change in spin polarization will cause a non zero spin flux into layer C,
i.e.\ the spin current entering layer C carries a different spin than 
the spin current exiting layer C. 
The net spin flux into layer C acts as a torque on 
the magnetization of layer C. 
In a SS system a current along the spiral axis  passes
through atomic thick layers with non parallel magnetization.
The spin of the electronic states that carry the current 
rotates coherently with the local
magnetization of the spiral.\cite{OH}
The precessing motion of 
the spins induces a torque on the magnetization that tends to rotate
the spin spiral. In the ground state 
the net STT from all the electronic states cancels out since
there is an equal number of 
electronic states with a current parallel and anti parallel 
to the spiral axis. 
But if a net current flows along the spiral axis 
no cancellation will occur and
a STT will act on 
the total angular momentum of the atoms in the material.
The torque will cause the 
magnetization to rotate with a frequency proportional 
to the applied current,
or equivalently make 
the SS slide along the spiral axis. 

In order to quantify this effect
we calculate the STT from a method similar to the method 
used in Ref. \onlinecite{Stiles} for multilayer systems. 
The calculations 
are based on calculations of the spin flux using the spin current density
tensor ${\bf Q}$ (and the STT is investigated both for model systems
and by using first principles calculations of real systems). ${\bf Q}$
is given by
%
%
\begin{equation}
{\bf Q}_{n{\bf k}}({\bf r})=\mathrm{Tr \, Re} \left\{\psi^\dagger_{n{\bf k}}({\bf r})\,
{\bf S} \otimes\hat{{\bf v}}\psi_{n{\bf k}}({\bf r})\right\},
\end{equation}
where ${\bf S}$ is the spin operator, $\hat{{\bf v}}$ is the velocity 
operator and 
$\psi_{n{\bf k}}$
are the spinor wave functions with band and 
wave vector indexes $n$ and ${\bf k}$ respectively.
The torque exerted by an 
electronic state with band index $n$ and wave vector ${\bf k}$
on the angular momentum within a volume $V$ enclosed 
by the surface $S$  is given by its 
spin flux into $V$,
\begin{equation}
\int_S{\bf Q}_{n{\bf k}}\cdot d{\bf S}=-{\partial {\bf J}_{n{\bf k}}
\over \partial t}.
\end{equation}
In the above equation the torque is expressed as the change of total 
angular momentum $\partial {\bf J}_{n{\bf k}}/ \partial t$, within volume $V$.
%
%
%

The volume of interest is arbitrary, for 
the STT in multilayer systems one considers the torque acting on a magnetic 
layer. 
For SS systems the STT on individual atoms
can be calculated using non overlapping spheres centered at the atoms.
The total torque on an atom is obtained by summing the torque from all the 
occupied electron states. If the system is in its equilibrium ground state
the sum of all torques will be zero due to time invariance. 
However, if an external electric field ${\bf E}$ is applied 
the occupation of the states at 
the Fermi surface (FS) will change and the total torque on an atom can be 
calculated from a FS integral derived from semi classical 
Boltzmann linear response theory,
\begin{equation}
{\partial {\bf J}\over \partial t}=
{V_C\over(2\pi)^3}{\tau e\over \hbar}\sum_{n}\int_{FS}
{\partial {\bf J}_{n{\bf k}}\over \partial t}
\left(\nabla_{\bf k}\epsilon_{n{\bf k}}
\cdot{\bf E}\right){dS_{n \bf k}\over|\nabla_{\bf k}\epsilon_{n{\bf k}}|} \, ,
\label{Boltzman}
\end{equation}
where $V_C$, $\tau$, $e$ and $\epsilon_{n{\bf k}}$ 
are the volume of the system, electron relaxation 
time, electron charge, and band energies respectively.
The sum is over all bands crossing the Fermi level. The above equation 
defines a linear relation between the torque and the external field,  
\begin{equation}
{\partial {\bf J}\over \partial t}=\tau \sum_{n} \textsf{A}_n{\bf E}.
\label{LRJ}
\end{equation}
A similar 
expression is obtained for the resistivity, relating the current density with
the external field.
\begin{eqnarray}
{\bf j}&&=\tau \sum_n 
\textsf{B}
_n{\bf E}\label{LRI}
\label{curr}\\
&&={-1\over(2\pi)^3}{\tau e\over \hbar^2}\sum_{n}\int_{FS}
\nabla_{\bf k}\epsilon_{n{\bf k}}
\left(\nabla_{\bf k}\epsilon_{n{\bf k}}
\cdot{\bf E}\right){dS_{n \bf k}\over|\nabla_{\bf k}\epsilon_{n{\bf k}}|}.
\nonumber
\end{eqnarray}
%
Combining these two equations gives a linear relation between the torque
and the current density where the unknown electron relaxation time $\tau$ 
has been canceled,
\begin{equation}
{\partial {\bf J}\over \partial t}= (\sum_n\textsf{A}
_n)(\sum_m \textsf{B}
_m)^{-1}{\bf j}=
\textsf{C}
{\bf j}.
\label{LRJI}
\end{equation}
The above equation for the torque current tensor 
$\textsf{C}$
will now be evaluated for a real SS system, 
the helical SDW in the RE metal Er.
Erbium has a complex non-collinear magnetic structure which is strongly 
temperature dependent. Bulk Er has an hcp structure with c=5.585 {\AA} 
and a=3.56 {\AA}.\cite{Mackintosh} There is a rich variety of non-collinear 
magnetic structures and ordering vectors over different temperature ranges. 
Below 20K it has a conical SS, between 53.5 and 85K a longitudinal SDW and 
between 20K and 53.5K  is there an intermediate magnetic structure
(see Ref. \onlinecite{Taylor}). 

Although the formalism is valid for conical SS, in this work we 
will focus on planar spin spirals.
All the material specific quantities of Er used in the calculation of the 
matrix 
$\textsf{C}$
were calculated from first 
principle density functional theory.
The calculation of Er was made using the full-potential 
augmented plane wave plus local orbitals (FP-APW+lo) method as described in 
Ref. \onlinecite{Bettan}. The local spin density approximation (LSDA) 
as parametrized 
by von Barth and Hedin was used without use of any shape-approximation 
to the non-collinear magnetization, i.e.\ charge and magnetization 
densities as well as their conjugate potentials are allowed to vary freely 
in space both regarding magnitude and direction. A set of 248 k-points 
was used for converging the electron density. The SS was treated 
using the generalised Bloch theorem. The 4f-electrons were treated as core 
electrons. A large set of SS wave vectors $\bf{q}$ along 
he out-of-plane axis in the 
hcp lattice were calculated and an energy minimum was found for 
$q$=0.20 $2\pi/c$. 
The results are in agreement with previous calculations made for 
helical SDW in RE.\cite{Nord}

The spin current density tensor was calculated at the surface of 
the augmentation spheres of the atoms where
the APW  expansion can be written as a sum of plane waves, 
%
%
\begin{equation}
\psi_{n{\bf k}}({\bf r})=\sum_{\bf G}
\Big (\alpha a_{n{\bf k},\bf G}e^{i({\bf G}+{\bf k}-{\bf q}/2){\bf r}}+
\beta b_{n{\bf k},\bf G}e^{i({\bf G}+{\bf k}+{\bf q}/2){\bf r}}\Big )
\end{equation}
where $\alpha$ and $\beta$ are the up and down spinors respectively
and $\bf G$ are
the reciprocal lattice vectors. The plane wave coefficients $a$ and $b$ are
obtained from the first principles calculation. The spin flux into a 
sphere with radius $R$ 
centered at an atom at cite ${\bf r}_n$ is for plane waves given by the 
expression
\begin{widetext}
%
\begin{eqnarray}
\int_{S_{atom}} {\bf Q}_{n{\bf k}}\cdot d{\bf S}=&&{\hbar R^2\over m}\mathrm{Re}\,
\sum_{{\bf G},{\bf G'}}-i4\pi\Big[ \nonumber\\
&&\alpha^\dagger{\bf s}\alpha \; a^*_{n{\bf k},\bf G} 
 a_{n{\bf k},\bf G'} e^{-i({\bf G}-{\bf G'}){\bf r}_n}j_1(|{\bf G}-{\bf G'}|R)\,
 ({\bf G'}+{\bf k}-{\bf q}/2){\cdot} ({\widehat{{\bf G}-{\bf G'}}})\nonumber\\
+&& \alpha^\dagger{\bf s}\beta\;a^*_{n{\bf k},\bf G} 
 b_{n{\bf k},\bf G'} e^{-i({\bf G}-{\bf G'}-{\bf q}){\bf r}_n}j_1(|{\bf G}-{\bf G'}-{\bf q}|R)\,
({\bf G'}+{\bf k}+{\bf q}/2){\bf \cdot} ({\widehat{{\bf G}-{\bf G'}-{\bf q}}})\nonumber\\
+&&\beta^\dagger{\bf s}\alpha\;b^*_{n{\bf k},\bf G} 
 a_{n{\bf k},\bf G'} e^{-i({\bf G}-{\bf G'}+{\bf q}){\bf r}_n}j_1(|{\bf G}-{\bf G'}+{\bf q}|R)\, 
 ({\bf G'}+{\bf k}-{\bf q}/2){\cdot} ({\widehat{{\bf G}-{\bf G'}+{\bf q}}})\nonumber\\
+&&\beta^\dagger{\bf s}\beta\;b^*_{n{\bf k},\bf G}
 b_{n{\bf k},\bf G'} e^{-i({\bf G}-{\bf G'}){\bf r}_n}j_1(|{\bf G}-{\bf G'}|R)\,
({\bf G'}+{\bf k}+{\bf q}/2){\cdot} ({\widehat{{\bf G}-{\bf G'}}})\Big]
\end{eqnarray}
\end{widetext}
where $m$ is the electron mass and $j_1$ is the first spherical Bessel 
function. In order to get a correct description of the FS used in 
Eqs. (\ref{Boltzman}) and (\ref{curr}) a
$41\times 41\times 41$ k-point mesh was used to cover the first 
Brillouin zone.
 
The torque current matrix 
$\textsf{C}$
is evaluated for an 
Er atom situated at a site with magnetization direction $[1\, 0\, 0\,]$ and 
SS wave vector $[\,0\, 0\, q\,]$,
\[ \textsf{C}
=\hbar\left( \begin{array}{ccc}

   0 &   0 &   0\\
   0 &   0 &   0.5\\
   0 &   0 &   0\\
%
%

\end{array} \right) [\textrm{\AA}^2].\]
%
%
From the structure of the 
$\textsf{C}$
matrix we 
conclude that the torque induced by a current along the spiral axis
causes the SS to translate
along the spiral axis, which is equivalent to a rigid rotation of the spiral. 
In order to calculate the rotation frequency as a function of the current
density we also need to consider the total angular momentum 
of the Er atoms
$J=15/2$ given
by Hund's rules as the sum of orbital and spin angular momentum 
$L=6$ and $S=3/2$ 
. 
The torque calculated using the 
$\textsf{C}$
matrix will cause the total angular 
momentum of the 
atom as well as the SS to precess with 0.07 GHz if a current 
of $10^7 \textrm{A/cm}^2$ flows along the spiral axis of bulk Er.

\begin{figure}
\includegraphics[scale=0.4]{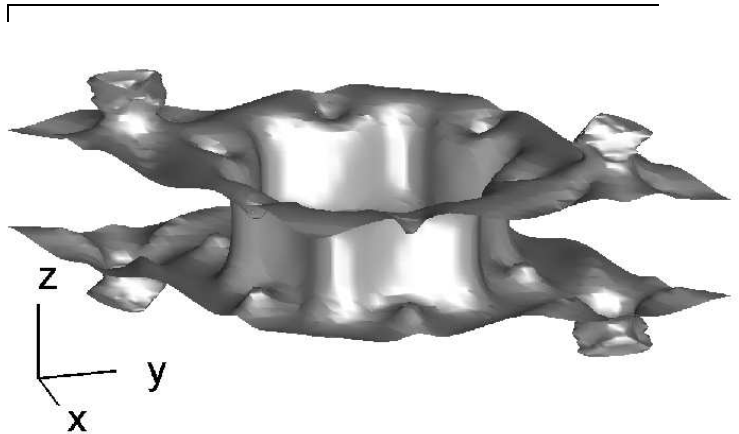}%
\caption{\label{FS}The Fermi surface of Er that contributes the most to 
electron transport along the SS axis. The spiral axis is along the z 
direction.}
\end{figure}

\begin{figure}
\includegraphics[scale=0.6]{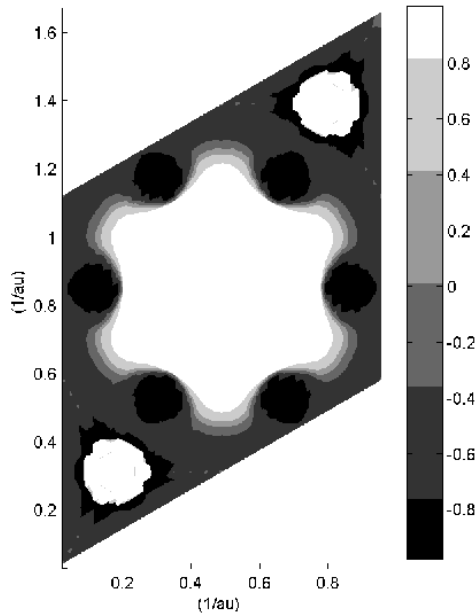}%
\caption{\label{FSSP}Projection of the Fermi surface of Fig. (\ref{FS}) on the 
x-y plane. The spin polarization parallel with the local magnetization 
direction $P$ is
given by the colour code, 1 means that the spin polarization is parallel 
with the local magnetization direction, for details see text.}
\end{figure}

By analyzing the quantities in Eq. (\ref{LRJI}) one finds that the main 
contribution to the 
$\textsf{C}$
matrix comes from the 
the band whose FS is shown in Fig. \ref{FS}.
This surface is the remains of the FS which
has the nesting features that drives some of the RE solids to have a 
helical SDW.\cite{Nord}
The large contribution from this surface 
is due to that it has a large area being perpendicular
to the spiral axis that contributes to the electron transport along 
the axis.
An estimate of the STT induced by the electronic states on this FS 
can be obtained from the following considerations.
As the conduction electrons flow along the spiral axis the component of their 
spins that are perpendicular to the SS axis i.e.\ for planar SS parallel to the local magnetization direction
%
will be precessing around the SS axis.\cite{OH} The size of the parallel
component $P$ can be estimated by the expression
\begin{equation}
P({\bf k})=\left< {\widehat{{\bf m}}({\bf r})\cdot{\bf s}_{\bf k}({\bf r})
\over |{\bf s}_{\bf k}({\bf r})|}\right>,\label{Pol}
\end{equation}
with
\begin{eqnarray}
{\bf s}_{\bf k}({\bf r})=\psi^\dagger_{\bf k}({\bf r}){\bf
  s}\psi_{\bf k}({\bf r}), \nonumber\\
\widehat{{\bf m}}({\bf r})=[cos({\bf qr}),sin({\bf qr}),0], \nonumber
\end{eqnarray}
where $\widehat{{\bf m}}({\bf r})$
is the local magnetization and $\langle...\rangle$ means space average.
$P$ is a measure of the local spin polarization and the value of $P$ for
the states at the FS in Fig. \ref{FS}
is shown in Fig. \ref{FSSP}. In Fig. \ref{FSSP} is the 
local spin polarization $P$ calculated using the interstitial region 
between two atomic planes i.e.\ the space average in Eq. (\ref{Pol}) 
is not done over the whole unit cell.
%
From Fig. \ref{FSSP} we estimate that the spin of the electron states at the FS
on average are tilted
$30^\circ$ from the spiral axis in the opposite direction of 
the local magnetization since the average spin polarization 
$P$ for these states is $-0.5$.
The current carried by an electronic state 
transfers spin upon passing an atomic layer perpendicular to the SS axis.
This is since its parallel spin component will rotate with $q\pi$ [rad] for each
layer the current passes. For states on the FS a spin of 
$(\hbar/2)\pi P q$ [Js] will be transfered per layer.
A current of 1[A/m$^2$] along 
the spiral
axis induces a STT of $(\hbar/2)2\pi PqA/e$ 
[J] per unit cell, where $e$ is the
electron charge and $A=a^2\sqrt{3}/2$ is the cell area.
The STT induced by 1A/m$^2$ along the spiral axis causes the SS to rotate with
$(PqA)/(4Je)$ [Hz]. This estimate of the
rotation frequency gives four times the value obtained from the 
$\textsf{C}$
matrix but catches the order of the effect.

For STT in trilayer systems there is a critical current density 
required to switch the magnetization which magnitude is 
governed by magnetic anisotropies.\cite{Stiles2} 
For SS systems there will also be a critical current needed 
to overcome the energy barrier due to magnetic anisotropies.
This implies
that the spiral rotation frequency will scale linearly with the current along
the spiral axis for currents larger than some critical current 
of the system. But since the STT in SS systems is a bulk 
phenomenon the critical 
current could be smaller then the critical current for trilayer systems.
Moreover, for the RE systems the anisotropies can be manipulated 
to a large degree, by varying the RE species by alloying, while
keeping the helical SDW magnetic order.\cite{Mackintosh}

\begin{figure}
\includegraphics[scale=0.5]{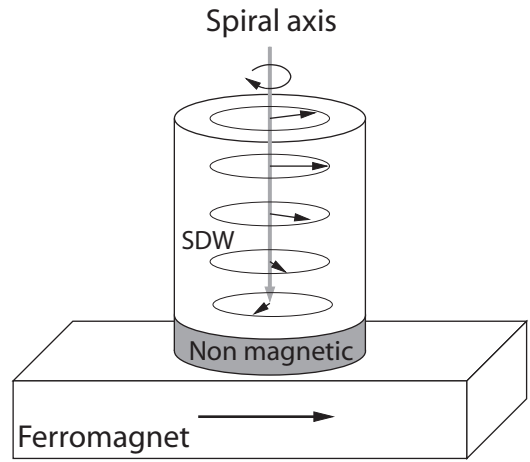}%
\caption{\label{Pillar} Device for experimental verification of STT in 
a material with a helical SDW.}
\end{figure}

We suggest that the rotation frequency of the SS
as a function of current density can be experimentally verified in the following way.
Consider a nano pillar with a layer of a nonmagnetic material in between a SS layer,
with its spiral axis perpendicular to the 
interface, and a ferromagnetic material as shown in Fig. \ref{Pillar}. 
In principle all these three layers can be RE based, e.g.\ with Gd as the ferromagnet and Lu as the non-magnetic species.
This system will in similarity with 
GMR and TMR systems have a conductance perpendicular to the layers dependent 
on the relative magnetization directions of the interfaces.\cite{GMR1,TMR1}
If the magnetization of the ferromagnetic material 
is fixed and the magnetization of the SS layer precesses then the 
conductivity perpendicular to the layers will be time dependent. The oscillating
conductivity can be used as a probe of the rotation frequency of the SS. 

\begin{acknowledgments}
We acknowledge support from the Swedish Research Council (VR) and the 
Foundation for Swedish Strategic Research (SSF). Parts of the calculations 
has been performed at the National Supercomputing Center NSC in Link{\"o}ping.
\end{acknowledgments}
\bibliography{stt}

\end{document}